\documentclass[preprint,groupedaddress,pre]{revtex4}

\input{tcilatex}

\begin{document}

\title{ Covariant EBK quantization of the electromagnetic two-body problem}
\author{Jayme De Luca}
\email[corresponding author; email address:]{ deluca@df.ufscar.br}
\affiliation{Universidade Federal de S\~{a}o Carlos, \\
Departamento de F\'{\i}sica\\
Rodovia Washington Luis, km 235\\
Caixa Postal 676, S\~{a}o Carlos, S\~{a}o Paulo 13565-905}
\date{\today }

\begin{abstract}
We discuss a method to transform the covariant Fokker action into an
implicit two-degree-of-freedom Hamiltonian for the electromagnetic two-body
problem with arbitrary masses. This dynamical system appeared 100 years ago
and it was popularized in the 1940's by the still incomplete Wheeler and
Feynman program to quantize it as a means to overcome the divergencies of
perturbative QED. Our finite-dimensional implicit Hamiltonian is closed and
involves no series expansions. The Hamiltonian formalism is then used to
motivate an EBK quantization based on the classical trajectories with a
non-perturbative formula that predicts energies free of infinities.
\end{abstract}

\maketitle

\section{Introduction}

The usual Hamiltonian description of two-body dynamics is surprisingly
restrictive within relativity physics: If Lorentz transformations are to be
represented by canonical transformations, only non-interacting two-body
motion can be described. This is the content of the no-interaction theorem
of 1964, which later in 1984 was proved for the usual Lagrangian description
as well \cite{nointeract}. A covariant version of Hamiltonian dynamics,
constraint dynamics, was invented to overcome this group-theoretical
obstacle, but this too has limited applicability \cite{Comar}. In this work
we start from a relativistic physical theory: the time-symmetric
relativistic action-at-a-distance electrodynamics, for which no constraint
description exists at present. We discuss the passage from the covariant
Fokker Lagrangian to an implicit Hamiltonian formalism defined in closed
form without power expansions. We derive a two-degree of freedom Hamiltonian
for the electromagnetic two-body problem \cite{two} with arbitrary masses
and with either repulsive or attractive interaction. In our description a
Lorentz transformation is represented by a canonical transformation followed
by a rescaling of the evolution parameter. We use this covariant Hamiltonian
formalism to motivate an EBK quantization of the electromagnetic two-body
problem that uses the classical trajectories to approximate the quantum
energies with a formula that is non-perturbative and free of infinite
quantities.

In 1903, Schwarzchild proposed a relativistic interaction between charges
that was time reversible precisely because it involved retarded and advanced
interactions symmetrically \cite{Schwarz}. The same model reappeared in the
1920s in the work of Tetrode and Fokker \cite{Tetrode-Fokker} and it finally
became electromagnetic theory after Wheeler and Feynman showed that this
direct-interaction theory can describe all the classical electromagnetic
phenomena (i.e. the classical laws of Coulomb, Faraday, Amp\`{e}re, and
Biot-Savart) \cite{Fey-Whe,Leiter}. In particular, Wheeler and Feynman
showed in 1945 that in the limit where the electron interacts with a
completely absorbing universe, the response of this universe to the
electron's field is equivalent to the \emph{local} Lorentz-Dirac
self-interaction theory \cite{Dirac} without the need of mass
renormalization \cite{Fey-Whe}. Complete absorption is added to the theory
as an approximation to uncouple away from the detailed neutral-delay
dynamics of the other charges of the universe. For other approximations see
Ref. \cite{Narlikar}. \ The Wheeler and Feynman program \cite{Mehra} to
quantize the action-at-distance electrodynamics and overcome the infinities
of QED is still not implemented because of the lack of a Hamiltonian
description, which was a motivation for the present work.

The relativistic action-at-a-distance electrodynamics is better regarded as
a many-body electromagnetic theory, once it is based on a
parametrization-independent action involving two-body interactions only,
without the mediation by fields \cite{Fey-Whe,Leiter}. The isolated
electromagnetic two-body problem, away from the other charges of the
universe, is a time-reversible dynamical system defined by the action 
\begin{equation}
S_{F}=-\int m_{1}ds_{1}-\int m_{2}ds_{2}-e_{1}e_{2}\int \int \delta
(||x_{1}-x_{2}||^{2})\dot{x}_{1}\cdot \dot{x}_{2}ds_{1}ds_{2},
\label{Fokker}
\end{equation}%
where $x_{i}(s_{i})$ represents the four-position of particle $i=1,2$
parametrized by its arc-length $s_{i}\,$, double bars stand for the
four-vector modulus $||x_{1}-x_{2}||^{2}\equiv (x_{1}-x_{2})\cdot
(x_{1}-x_{2})$, and the dot indicates the Minkowski scalar product of
four-vectors with the metric tensor $g_{\mu \nu }$\ ($%
g_{00}=1,g_{11}=g_{22}=g_{33}=-1$) (the speed of light is $c=1$). The
integration in Eq. (\ref{Fokker}) can not be restricted to a segment of
trajectory going from an initial to a final time, because Lagrangian (\ref%
{Fokker}) at the end-points involves the future and past light-cones, which
are out of the segment. This difference from our Poincar\'{e}-invariant
Lagrangian to the usual Hamilton's principle of Galilei-invariant mechanics
demands that action (\ref{Fokker}) be defined by the whole orbit. A sensible
mathematical variation of Eq.(\ref{Fokker}) is along trajectory variations $%
\delta x_{i}(t)$\ that vanish at both $t=\pm \infty $, such that $\delta
S_{F}$ is a sensible finite quantity to be minimized. This difference is
discussed in \cite{Staruskiewicz}, and we shall henceforth ignore it and
derive our equations of motion \emph{formally} by extremizing action (\ref%
{Fokker}). The attractive two-body problem is defined by Eq.(\ref{Fokker})
with $e_{1}=-e_{2}\equiv e$ (Hydrogen atom), while the repulsive problem is
defined by Eq. (\ref{Fokker}) with $e_{1}=e_{2}\equiv e$. For the
electromagnetic two-body problem with arbitrary masses (Eq. (\ref{Fokker})),
the only known analytical solution is the circular orbit for the attractive
problem \cite{Schonberg,Schild}. A comprehensive discussion of literature on
the Fokker Lagrangian is given in Ref. \cite{two}. Some existence and
uniqueness results for this two-body system are reviewed in Appendix I.

This paper is divided as follows: In Section II we construct a Hamiltonian
that is general enough to describe the non-runaway solutions of the Fokker
Lagrangian for the two-body problem with arbitrary masses and with either
attractive or repulsive interaction. In Section III we use this Hamiltonian
in an EBK quantization procedure that is non-perturbative and can provide a
set of energies free of infinite quantities. In Section IV we put the
discussions and conclusion. In Appendix I we construct a Lorentz frame where
a generic non-runaway orbit of the Fokker Lagrangian is time-reversible (the
CMF frame).

\bigskip

\section{Outline of the Method}

The method to describe twice-monotonic orbits of the Fokker Lagrangian by an
implicit Hamiltonian formalism was discussed in Ref. \cite{two} for the
special case of equal masses with repulsive interaction. In the following we
generalize this result for the case of arbitrary masses with either
repulsive or attractive interaction.

The first step of the method consists of a transformation to new variables 
\cite{Staruskiewicz} 
\begin{eqnarray}
\xi _{1} &\equiv &t_{1}-x_{1},\qquad \zeta _{1}\equiv t_{1}+x_{1},
\label{defining} \\
\xi _{2} &\equiv &t_{2}-x_{2},\qquad \zeta _{2}\equiv t_{2}+x_{2}.  \nonumber
\end{eqnarray}%
This transformation splits the action integral (\ref{Fokker}) into the sum
of two separate actions 
\begin{equation}
S_{F}=\frac{1}{2}(S_{a}+S_{b}),  \label{splitFokk}
\end{equation}%
with

\begin{eqnarray}
S_{a} &=&-\int m_{1}(d\xi _{1}d\zeta _{1})^{1/2}-\int m_{2}(d\xi _{2}d\zeta
_{2})^{1/2}  \label{eqSa} \\
&&-e_{1}e_{2}\int \int \frac{\delta (\zeta _{1}-\zeta _{2})}{|\xi _{1}-\xi
_{2}|}(d\xi _{1}d\zeta _{2}+d\xi _{2}d\zeta _{1}),  \nonumber
\end{eqnarray}%
and 
\begin{eqnarray}
S_{b} &=&-\int m_{1}(d\xi _{1}d\zeta _{1})^{1/2}-\int m_{2}(d\xi _{2}d\zeta
_{2})^{1/2}  \label{eqSb} \\
&&-e_{1}e_{2}\int \int \frac{\delta (\xi _{1}-\xi _{2})}{|\zeta _{1}-\zeta
_{2}|}(d\xi _{1}d\zeta _{2}+d\xi _{2}d\zeta _{1}).  \nonumber
\end{eqnarray}%
Integration of the $\delta $ function in Eq. (\ref{eqSa}) gives a nonzero
contribution only where $\zeta _{1}$ and $\zeta _{2}$ take equal values $%
\zeta _{1}=\zeta _{2}$ =$\zeta $, and this $\zeta $ is the natural evolution
parameter of action $S_{a}$ (henceforth called type $a$ foliation).
Analogously, for Eq. (\ref{eqSb}), integration over the $\delta $ function
produces a nonzero contribution only where $\xi _{1}=\xi _{2}=\xi $, and
this $\xi $ is the natural time parameter of action $S_{b}$ (henceforth
called type $b$ foliation). Notice that with type $a$ foliation particles
are in the light-cone condition $(x_{1}-x_{2})^{2}-(t_{1}-t_{2})^{2}=0$ and
particle $2$ is ahead of particle $1$ in time by $r_{a}$. The light-cone
distance $r_{a}$ in the type $a$ foliation is%
\begin{equation}
r_{a}=-\frac{1}{2}\bigskip (\xi _{1}-\xi _{2}).  \label{definera}
\end{equation}%
With type $b$ foliation particles are also in the light-cone condition, and
particle $2$ is behind in time by $r_{b}$ , and the light-cone distance $%
r_{b}$ is 
\begin{equation}
r_{b}=\frac{1}{2}\bigskip (\zeta _{1}-\zeta _{2}).  \label{definerb}
\end{equation}%
After expressing action (\ref{eqSa}) in terms of the time-like parameter $%
\zeta $, the associated Euler-Lagrange equation is a simple ordinary
differential equation. The Euler-Lagrange problem for action (\ref{eqSb}) is
analogous, with $\zeta $ replaced by $\xi .$ A Lagrangian (such as Eqs. (\ref%
{eqSa}) and (\ref{eqSb})), whose Euler-Lagrange equation is an ordinary
differential equation is henceforth called to be in the local form. Notice
that Lagrangians (\ref{eqSa}) and (\ref{eqSb}) are also separately covariant.

The fact that the Fokker Lagrangian is not in the local form is a motivation
to search for an equivalent covariant and local Lagrangian, and the natural
candidates would be $S_{a}$ and $S_{b}$ of Eqs. (\ref{eqSa}) and (\ref{eqSb}%
). With\ that in mind, one could speculate if an orbit that extremizes Eq. (%
\ref{Fokker}) could also extremize actions $S_{a}$ (Eq. (\ref{eqSa})) $%
(\delta S_{a}=0)$ and $S_{b}$ (Eq. (\ref{eqSb})) ($\delta S_{b}=0)$
separately. Each separate minimization, in general, yields a different
trajectory, and as discussed in \cite{two},\emph{\ }these trajectories are
never equal. In Ref. \cite{two} we found that the trajectories of the
following alternative pair of Lagrangians%
\begin{equation}
\delta S_{a}=\delta G,  \label{var1}
\end{equation}%
and 
\begin{equation}
\delta S_{b}=-\delta G,  \label{var2}
\end{equation}%
coincide for a choice of the Lagrangian $G$. A trajectory that satisfies
Eqs. ($\ref{var1}$) and ($\ref{var2}$) will also extremize the Fokker action
(\ref{Fokker}), a simple consequence of Eqs. (\ref{splitFokk}), (\ref{var1})
and (\ref{var2}):%
\begin{equation}
\delta S_{F}=\frac{1}{2}\delta S_{a}+\frac{1}{2}\delta S_{b}=\frac{1}{2}%
\delta G-\frac{1}{2}\delta G=0.  \label{consequence}
\end{equation}%
Our problem is then to fix the Lagrangian $G$ \ such that Eqs. ($\ref{var1}$%
) and ($\ref{var2}$) yield the same trajectory and such that Eqs. (\ref{var1}%
) and (\ref{var2}) be ordinary differential equations (rather than delay
equations). Such $G$ is henceforth called a bilocal ghost Lagrangian.

In Appendix I we prove that any non-runaway orbit of (\ref{Fokker}) \ has a
Lorentz frame in which it is time-reversible (henceforth called the CMF
frame). In the construction that follows we operate on this CMF Lorentz
frame. Since the orbit is time reversible, we can restrict the minimization
of Eqs. (\ref{var1}) and (\ref{var2}) to the set of \ time-reversible orbits
([$x_{1}(-t)=x_{1}(t)$] and [$x_{2}(-t)=x_{2}(t)$ ]). We also add the
physical property that both advanced and retarded distances decrease
monotonically to a point of minimum and then start increasing monotonically
again. We henceforth call this set of orbits the CMF set. The piecewise
monotonic property is a consequence of the velocity being a monotonic
function of time, which was proved in Appendix 1 of Ref. \cite{two} for the
repulsive case. The generalization to the attractive case along symmetric
orbits is trivial. We henceforth refer to a CMF orbit simply as a
twice-monotonic orbit. \ Assuming that a twice-monotonic and time-reversible
orbit exists in our given Lorentz frame, as this orbit extremizes Eq. (\ref%
{Fokker}) over all orbits, it is necessarily an extremum inside the CMF set.

In the following we generalize some integral identities first used in Ref. 
\cite{two} for CMF orbits. The time-reversal operation maps type $a$
parametrization onto type $b$ and acts on CMF orbits by the following map: $%
\zeta _{1,2}\rightarrow $ $-\xi _{1,2}$ , $\xi _{1,2}\rightarrow -\zeta
_{1,2}$, $r_{a}\rightarrow r_{b}$. The simplest type of integral identity,
valid for an arbitrary integrable function $\phi (x)$ of the real variable,
is%
\begin{equation}
\int_{a}\phi (r_{a})d\zeta =\int_{b}\phi (r_{b})d\xi .  \label{defphi}
\end{equation}%
The lower index of the integral denotes the parametrization type, and Eq. (%
\ref{defphi}) is a consequence of the coordinate transformation induced by
the time-reversal symmetry of the CMF set $\zeta \rightarrow -\xi $ (the
Jacobian of this transformation is one). \ For the attractive problem we can
restrict the integration to twice the finite interval between two
consecutive collisions (the period), while for the repulsive problem we must
define the above integrals over the infinite interval. \ In this work we
ignore questions of convergence of the integrals, an ambiguity inherited
from the Wheeler-Feynman theory and discussed in Ref. \cite{Staruskiewicz}.
The same time-reversal action ($\zeta _{1,2}\rightarrow -\xi _{1,2},\xi
_{1,2}\rightarrow -\zeta _{1,2}$) along CMF orbits generates the following
identities for arbitrary functions $V_{1}(r)$, $V_{2}(r)$, $\alpha _{1}(r)$
and $\alpha _{2}(r)$ of the real variable

\begin{eqnarray}
\int_{a}V_{1,2}(r_{a})(\frac{d\xi _{1}}{d\zeta }+\frac{d\xi _{2}}{d\zeta }%
)d\zeta &=&\!\int_{b}V_{1,2}(r_{b})(\frac{d\zeta _{1}}{d\xi }+\frac{d\zeta
_{2}}{d\xi })d\xi ,  \label{defValphabeta} \\
\int_{a}\alpha _{1,2}(r_{a})(d\xi _{1}d\zeta _{1})^{1/2} &=&\int_{b}\alpha
_{1,2}(r_{b})(d\xi _{1}d\zeta _{1})^{1/2}.
\end{eqnarray}%
The above identities suggest that we use a ghost Lagrangian $G$ of type%
\begin{equation}
G=\int_{a}[\phi (r_{a})+\frac{1}{2}V_{1}(r_{a})\dot{\xi}_{1}+\frac{1}{2}%
V_{2}(r_{a})\dot{\xi}_{2}+\alpha _{1}(r_{a})\sqrt{\dot{\xi}_{1}}+\alpha
_{2}(r_{a})\sqrt{\dot{\xi}_{2}}]d\zeta .  \label{generalGa}
\end{equation}%
The dot over $\xi _{1,2}$ in Eq. (\ref{generalGa}) indicates derivative
respect to $\zeta $ ( the time-parameter of case $a$). This $G$ is in the
local form when added to $S_{a\text{ }}$, where $\zeta $ plays the role of
the time parameter and the coordinates are $\xi _{1}$, $\xi _{2}$. When this
same $G$ is subtracted from action $S_{b}$, the integral identities allow us
to express $G$ as 
\begin{equation}
G=\int_{b}[\phi (r_{b})+\frac{1}{2}V_{1}(r_{b})\dot{\zeta}_{1}+\frac{1}{2}%
V_{2}(r_{b})\dot{\zeta}_{2}+\alpha _{1}(r_{b})\sqrt{\dot{\zeta}_{1}}+\alpha
_{2}(r_{b})\sqrt{\dot{\zeta}_{2}}]d\xi ,  \label{generalGb}
\end{equation}%
which is also in the local form for action $S_{b}$, with $\xi $ being the
time parameter and the coordinates being $\zeta _{1}$ and $\zeta _{2}.$
Notice that the dot over $\zeta _{1,2}$ in equation (\ref{generalGb})
indicates derivative respect to $\xi $ (the time-parameter of case $b$).
Lagrangian (\ref{generalGa}) is the most general ghost Lagrangian whose
associated Hamiltonian involves quadratic rational functions of the momenta,
and we shall see that it suffices to describe any non-runaway orbit with two
monotonic branches. \ 

The condition $\dot{r}=0$ divides the phase space of a twice-monotonic orbit
in two disjoint regions according to whether $\dot{r}>0$ or $\dot{r}<0$ . In
each disjoint region the Lagrangian has a branch that can be uniquely
inverted to a Hamiltonian formalism. To avoid the overloaded notation of
Ref. \cite{two}, we henceforth drop all indications about branches. Notice
that after inclusion of the ghost Lagrangian, the problem $\delta
S_{a}=\delta G$ \ of Eq. (\ref{var1}) implies the Euler-Lagrange equations
for $L_{a}=S_{a}-G$

\begin{equation}
L_{a}=-\int [M_{1a}\sqrt{\dot{\xi}_{1}}+M_{2a}\sqrt{\dot{\xi}_{2}}+\frac{%
e_{1}e_{2}}{|\xi _{1}-\xi _{2}|}+\frac{1}{2}V_{1}(r_{a})\dot{\xi}_{1}+\frac{1%
}{2}V_{2}(r_{a})\dot{\xi}_{2}+\phi (r_{a})]d\zeta ,  \label{Lagrangian-a}
\end{equation}%
where $M_{1a}\equiv m_{1}+\alpha _{1}(r_{a})$ and $M_{2a}\equiv m_{2}+\alpha
_{2}(r_{a})$.\ The time-reversed problem $\delta S_{b}=-\delta G$ is
described by $L_{b}=S_{b}+G$,

\begin{equation}
L_{b}=-\int [M_{1b}\sqrt{\dot{\zeta}_{1}}+M_{2b}\sqrt{\dot{\zeta}_{2}}+\frac{%
e_{1}e_{2}}{|\zeta _{1}-\zeta _{2}|}-\frac{1}{2}V_{1}(r_{b})\dot{\zeta}_{1}-%
\frac{1}{2}V_{2}(r_{b})\dot{\zeta}_{2}-\phi (r_{b})]d\xi ,
\label{Lagrangian-b}
\end{equation}%
with $M_{1b}\equiv m_{1}-\alpha _{1}(r_{b})$ and $M_{2b}\equiv m_{2}-\alpha
_{2}(r_{b})$ . The Hamiltonian in each case and branch is given by 
\begin{equation}
H_{a}=\frac{-1}{4}\{\frac{M_{1a}^{2}}{(p_{1}+\frac{1}{2}V_{1}+\frac{%
e_{1}e_{2}}{|\xi _{1}-\xi _{2}|})}+\frac{M_{2a}^{2}}{(p_{2}+\frac{1}{2}V_{2}+%
\frac{e_{1}e_{2}}{|\xi _{1}-\xi _{2}|})}\}-\phi (r_{a}),  \label{hamia}
\end{equation}%
and 
\begin{equation}
H_{b}=\frac{-1}{4}\{\frac{M_{1b}^{2}}{(p_{1}-\frac{1}{2}V_{1}+\frac{%
e_{1}e_{2}}{|\zeta _{1}-\zeta _{2}|})}+\frac{M_{2b}^{2}}{(p_{2}-\frac{1}{2}%
V_{2}+\frac{e_{1}e_{2}}{|\zeta _{1}-\zeta _{2}|})}\}+\phi (r_{b}).
\label{hamib}
\end{equation}%
Notice that the Hamiltonian $H_{a}$ depends only on $r_{a}=-\frac{1}{2}(\xi
_{1}-\xi _{2})$, such that $P_{a}=p_{1}+p_{2}$ is a constant of motion.
Analogously for type $b$ parametrization, Hamiltonian $H_{b}$ depends only
on $r_{b}=\frac{1}{2}(\zeta _{1}-\zeta _{2})$, such that $P_{b}=p_{1}+p_{2}$
is a constant of motion. The constant $P_{a}=p_{1}+p_{2}$ of case $a$
suggests the following canonical change of variables 
\begin{eqnarray}
X &\equiv &\frac{1}{2}(\xi _{1}+\xi _{2}),\quad P\equiv p_{1}+p_{2}
\label{canonical} \\
x &\equiv &\frac{1}{2}(\xi _{1}-\xi _{2}),\quad p=p_{1}-p_{2}.  \nonumber
\end{eqnarray}%
For type $b$ we use the analogous transformation with $\xi $ replaced by $%
\zeta $ in Eq. (\ref{canonical}). We use Eq. (\ref{canonical}) to express $%
p_{1}$ and $p_{2}$ of Eq. $(\ref{hamia})$ in terms of $P$ and $p$ and
substitute into the condition $H_{a}(r_{a},p_{a},E,P)=E_{a}$ , yielding a
quadratic equation for $p$, with solutions%
\begin{equation}
p_{a}(r_{a},E_{a},P_{a})=\frac{V_{2}-V_{1}}{2}+\frac{\Delta _{a}}{%
(E_{a}+\phi )}\mp \sqrt{(P_{a}+\frac{Q_{a}}{(E_{a}+\phi )}+\frac{1}{2}V_{1}+%
\frac{1}{2}V_{2}+\frac{e_{1}e_{2}}{r_{a}})^{2}+(\frac{\Delta
_{a}^{2}-Q_{a}^{2}}{(E_{a}+\phi )^{2}})},  \label{Coulomb}
\end{equation}%
where $Q_{a}\equiv \frac{1}{4}(M_{1a}^{2}+M_{2a}^{2})$ , $\Delta _{a}\equiv 
\frac{1}{4}(M_{2a}^{2}-M_{1a}^{2})$ , $r_{a}=|x|$ and the plus or minus
describes the two branches. The separation for case $b$ is analogous. Notice
that in Ref. \cite{two} we used only one function $V=\frac{1}{2}%
(V_{1}+V_{2}),$ a trivial gauge transformation if $V_{1}$ and $V_{2}$ are
functions only of $r_{a}$, but nontrivial if $V_{1}$ and $V_{2}$ depend on $%
E $ and $P$, as in the generalization that follows. This generalization is
necessary for the different mass case.

As we show below, it is always possible to adjust the potentials such that
Hamiltonians (\ref{hamia}) and (\ref{hamib}) have a common trajectory. The
drawback is that these so adjusted potentials depend on $E$ and $P$ as well.
This dependence suggests that we generalize the potentials of Eqs. (\ref%
{hamia}) to functions of $E$ and $P$ \cite{two}. For example, the potential $%
\phi (r_{a})$ should be generalized to $\phi =\phi (r_{a},E_{a},P_{a})$ in
case $a$ and to $\phi =\phi (r_{b},E_{b},P_{b})$ in case $b$ (an analogous
generalization goes for $V_{1,2}$ and $\alpha _{1,2}$). The generalized
Hamiltonian is still a function of phase space, defined implicitly by Eqs. (%
\ref{hamia}) and (\ref{hamib}), because $E$ and $P$ are functions of the
phase space. This generalization must be done carefully though, because in
general the equations of motion of Hamiltonians (\ref{hamia}) and (\ref%
{hamib}) with implicitly defined potentials do not correspond to the
Lagrangian equations of (\ref{Lagrangian-a}) and (\ref{Lagrangian-b})
anymore, but rather to some highly involved Lagrangian with a complex
dependence on the velocities. On the other hand, simultaneous minimization
of a pair of Lagrangians with the simple form of (\ref{Lagrangian-a}) and (%
\ref{Lagrangian-b}) is very desirable, because the Wheeler-Feynman equation
of motion ($\delta S_{F}=0$ ) is then a consequence of Eq. (\ref{consequence}%
). In the following we construct a pair of \emph{auxiliary} Lagrangians with
the simple form (\ref{Lagrangian-a}) and (\ref{Lagrangian-b}) for any given
orbit. These \emph{auxiliary }Lagrangians carry different potential
functions for each orbit. \ All we need to know about these \emph{auxiliary }%
Lagrangians is their existence, as it guarantees that our orbit satisfies
the Wheeler-Feynman equation of motion by means of Eq. (\ref{consequence}).

\bigskip\ The \emph{auxiliary} Lagrangian is constructed from the potential
functions of $(r,P,E)$ as follows: For an orbit of energy $E_{o}$ and
momentum $P_{o}$, we define auxiliary \emph{fixed form }potentials: $\phi
^{\ast }(r)=\phi (r,E_{o},P_{o})$; $V_{1,2}^{\ast
}(r)=V_{1,2}(r,E_{o},P_{o}) $; $\alpha _{1,2}^{\ast }(r)=\alpha
_{1,2}(r,E_{o},P_{o})$, which in turn define \emph{auxiliary} Lagrangians by
Eqs. (\ref{Lagrangian-a}) and (\ref{Lagrangian-b}). The orbit of the \emph{%
auxiliary} Lagrangian (\ref{Lagrangian-a}) satisfies the \emph{auxiliary}
Hamiltonian equation of motion defined by Eq. (\ref{hamia}) with the \emph{%
fixed form} potentials. This equation is in general different from the
Hamiltonian equation of Eq. (\ref{hamia}) with the implicit potentials. To
fix this, we construct the potentials such that both equations are equal in
each case. For this it is sufficient to supplement the following conditions
in case $a$ 
\begin{eqnarray}
\frac{\partial H_{a}(p,P_{a},r,E_{a})}{\partial E_{a}} &=&0,
\label{firstwo-1} \\
\frac{\partial H_{a}(p,P_{a},r,E_{a})}{\partial P_{a}} &=&0,
\end{eqnarray}%
and analogously in case $b$ 
\begin{eqnarray}
\frac{\partial H_{b}(p,P_{b},r,E_{b})}{\partial E_{b}} &=&0,
\label{firstwo-2} \\
\frac{\partial H_{b}(p,P_{b},r,E_{a})}{\partial P} &=&0.
\end{eqnarray}%
These are four conditions imposed on the five potentials $\phi ,$ $V_{1,2}$
and $\alpha _{1,2}$. Before we proceed to construct potentials satisfying
the above conditions, it is convenient to find the explicit solution of
Hamiltonian (\ref{hamia}), which is accomplished by use of a canonical
transformation with a generating function $S$ given by 
\begin{equation}
S=PX+W(x,P,E)-E\zeta ,  \label{Jacobi}
\end{equation}%
where $W(x,P,E)$ is defined to satisfy the condition $p=\partial W/\partial
x $ , with $p$ given by Eq. (\ref{Coulomb}). The new momentum associated
with the old variable $X$ is the same old constant $P$ $=\partial S/\partial
X$ and the other new momentum is the energy $E$ (with this last definition
we exploit the fact that $E$ is already one argument of the implicit
potentials). We choose $S$ in the manner of Hamilton-Jacobi such that the
new Hamiltonian vanishes: $\ K=H+\frac{\partial S}{\partial \zeta }=0.$ As
the Hamiltonian is zero, the new coordinates are defined simply by two
constants $X_{0\text{ }}$and $C_{0}$

\begin{eqnarray}
X_{0} &=&\partial S/\partial P=X+\partial W/\partial P  \label{Hami-Jaco} \\
\quad C_{0} &=&-\partial S/\partial E=\zeta -\partial W/\partial E  \nonumber
\end{eqnarray}%
Eq. (\ref{Hami-Jaco}) defines $\zeta $ and $X$ as functions of $r_{a}\equiv
|x|$ for case $a$ (and $\xi $ and $X$ as a function of $r_{b}=|x|$ for case $%
b$ ). We shall also need the differential form of Eq. (\ref{Hami-Jaco})
relative to $x,$%
\begin{eqnarray}
dX &=&-(\partial ^{2}W/\partial x\partial P)dx=-(\partial p/\partial P)dx,
\label{contact} \\
d\zeta &=&(\partial W/\partial x\partial E)dx=(\partial p/\partial E)dx,
\label{contact2}
\end{eqnarray}%
where we have used $p=\partial W/\partial x$ (definition of the
Hamilton-Jacobi transformation) and exchanged the partial derivatives. The
differentials $dx_{i}$ and $dt_{i}$ along the trajectory are obtained using
Eq. (\ref{defining}) to relate particle coordinates to $X$ and $\zeta $
followed by use of \ Eqs. (\ref{contact}) and (\ref{contact2}) to relate $dX$
and $d\zeta $ to $dr_{a}=|dx|$ $.$ The explicit solution is 
\begin{eqnarray}
dt_{1a,b} &=&\pm \frac{1}{2}(\frac{\partial p_{a,b}}{\partial P}-\frac{%
\partial p_{a,b}}{\partial E}-1)dr_{a,b},  \label{tablea} \\
dt_{2a,b} &=&\pm \frac{1}{2}(\frac{\partial p_{a,b}}{\partial P}-\frac{%
\partial p_{a,b}}{\partial E}+1)dr_{a,b},  \nonumber \\
dx_{1a,b} &=&\frac{1}{2}(1-\frac{\partial p_{a,b}}{\partial P}-\frac{%
\partial p_{a,b}}{\partial E})dr_{a,b},  \nonumber \\
dx_{2a,b} &=&-\frac{1}{2}(\frac{\partial p_{a,b}}{\partial P}+\frac{\partial
p_{a,b}}{\partial E}+1)dr_{a,b},  \nonumber
\end{eqnarray}%
where the plus sign holds for case $a$ and the minus holds for case $b$ in
the time variables. Eq. (\ref{tablea}) gives the explicit solution in terms
of $p_{a,b}(r,E,P)$, as given by Eq. (\ref{Coulomb}).

As time reversal maps case $a$ into case $b$, \ a time-reversible orbit of
both (\ref{hamia}) and (\ref{hamib}) must satisfy $v_{1a}(r)=-v_{1b}(r)$ and 
$v_{2a}(r)=-v_{2b}(r)$ for $r\in \lbrack r_{o},\infty ].$ An economical way
to satisfy this condition is to express the functions $\partial p/\partial P$
and $\partial p/\partial E$ as

\begin{eqnarray}
\frac{\partial p_{a}}{\partial P} &=&\frac{\partial p_{b}}{\partial P}\equiv
-\frac{\cosh [s(r,E,P)]}{\sinh [s(r,E)]},  \label{condi2-1} \\
\frac{\partial p_{a}}{\partial E} &=&\frac{\partial p_{b}}{\partial E}\equiv 
\frac{F(r,E,P)}{\sinh [s(r,E,P)]},  \label{condi2-2}
\end{eqnarray}%
such that $v_{1a}(r)=-v_{1b}(r)$ and $v_{2a}(r)=-v_{2b}(r)$ \ are
automatically satisfied\ by use of Eqs. (\ref{tablea}). We are now ready to
obtain an analytical solution of Eqs. (\ref{firstwo-1}), (\ref{firstwo-2}), (%
\ref{condi2-1}) and (\ref{condi2-2}) in terms of the undetermined functions $%
s(r,E,P)$ and $F(s,E,P)$ defined by Eqs.(\ref{condi2-1}) and (\ref{condi2-2}%
). Our solution is defined such that $E_{a}=E_{b}\equiv E$ and $%
P_{a}=P_{b}\equiv P$. For case $a$ the equations of motion for $\xi _{1}$
and $\xi _{2}$ derived from Hamiltonian ( \ref{hamia}) with use of
conditions (\ref{firstwo-1}) are%
\begin{equation}
\frac{d\xi _{1}}{d\zeta }=\frac{\exp (s)}{F(r)}=\frac{(m_{1}+\alpha _{1})^{2}%
}{4(p_{1}+\frac{1}{2}V_{1}+\frac{e_{1}e_{2}}{2r_{a}})^{2}},  \label{Eq49}
\end{equation}%
and 
\begin{equation}
\frac{d\xi _{2}}{d\zeta }=\frac{\exp (-s)}{F(r)}=\frac{(m_{2}+\alpha
_{2})^{2}}{4(p_{2}+\frac{1}{2}V_{2}+\frac{e_{1}e_{2}}{2r_{a}})^{2}}.
\label{Eq50}
\end{equation}%
As discussed in Ref.\cite{two}, there are four possible square roots of Eqs.
(\ref{Eq49}) and (\ref{Eq50}), but only one choice does not pose a
constraint involving $s(r,E,P)$ and $F(r,E,P)$, leaving both functions
arbitrary. This physical choice of signs is defined by 
\begin{eqnarray}
P+p_{a}+V_{1}+\frac{e_{1}e_{2}}{r} &=&-\sqrt{F}(m_{1}+\alpha _{1})\exp
(-s/2),  \label{main1} \\
P-p_{a}+V_{2}+\frac{e_{1}e_{2}}{r} &=&\sqrt{F}(m_{2}+\alpha _{2})\exp (s/2).
\label{main2}
\end{eqnarray}%
where we have used $p_{1}=\frac{1}{2}(P_{a}+p_{a})$ and $p_{2}=\frac{1}{2}%
(P_{a}-p_{a})$ with $P_{a}=P.$ Case $b$ is analogous, and the motion of $%
\zeta _{1}$ and $\zeta _{2}$ derived from Hamiltonian ( \ref{hamib}) with
condition Eq. (\ref{firstwo-2}) defines the physical choice of the square
roots by 
\begin{eqnarray}
P+p_{b}-V_{1}+\frac{e_{1}e_{2}}{r} &=&\sqrt{F}(m_{1}-\alpha _{1})\exp (-s/2),
\label{main3} \\
P-p_{b}-V_{2}+\frac{e_{1}e_{2}}{r} &=&-\sqrt{F}(m_{2}-\alpha _{2})\exp (s/2).
\label{main4}
\end{eqnarray}%
We can use Eqs. (\ref{main1}) and (\ref{main2}) to eliminate $p_{1}$ and $%
p_{2}$ from Eq. (\ref{hamia}) with $E_{a}=E$, yielding

\begin{equation}
E+\phi =\frac{1}{2\sqrt{F}}[(m_{1}+\alpha _{1})\exp (s/2)-(m_{2}+\alpha
_{2})\exp (-s/2)],  \label{main5}
\end{equation}%
and an analogous use of Eqs. (\ref{main3}), (\ref{main4}) and (\ref{hamib})
with $E_{b}=E$ yields

\begin{equation}
E-\phi =-\frac{1}{2\sqrt{F}}[(m_{1}-\alpha _{1})\exp (s/2)-(m_{2}-\alpha
_{2})\exp (-s/2)].  \label{main6}
\end{equation}%
Notice that Eqs. (\ref{condi2-1}) and (\ref{condi2-2}) are immediately
satisfied if we assume 
\begin{equation}
p_{a}=p_{b}=p(E,P,r).  \label{main7}
\end{equation}%
Eqs. (\ref{condi2-1}) and (\ref{condi2-2}) actually allow $p_{a}$ and $p_{b}$
to differ by a\ trivial Gauge function $g(r)$ entering in Eq. (\ref{main7}),
but that can be absorbed in the definition of $V_{1}$ and $V_{2}$. We can
satisfy the seven conditions of Eqs. (\ref{main1})-(\ref{main7}) with
completely arbitrary functions $F(E,P,r)$ and $s(E,P,r)$ in the following
solution%
\begin{eqnarray}
\alpha _{1} &=&\frac{EF\exp (s/2)+(P+\frac{e_{1}e_{2}}{r})\exp (-s/2)}{\sqrt{%
F}\sinh (s)},  \label{solution} \\
\alpha _{2} &=&\frac{EF\exp (-s/2)+(P+\frac{e_{1}e_{2}}{r})\exp (s/2)}{\sqrt{%
F}\sinh (s)},  \nonumber \\
V_{1} &=&-m_{1}\sqrt{F}\exp (-s/2),  \nonumber \\
V_{2} &=&m_{2}\sqrt{F}\exp (s/2),  \nonumber \\
\phi &=&\frac{m_{1}\exp (s/2)-m_{2}\exp (-s/2)}{2\sqrt{F}}.  \nonumber
\end{eqnarray}%
and last with $p$ given by%
\begin{equation}
p=-\frac{[EF+(P+\frac{e_{1}e_{2}}{r})\cosh (s)]}{\sinh (s)}.  \label{psol}
\end{equation}%
Now that we have solved simultaneously all conditions (\ref{firstwo-1}), (%
\ref{firstwo-2}), (\ref{condi2-1}), (\ref{condi2-2}) and (\ref{main1})-(\ref%
{main6}), some remarks are in order:

(i) To see the generality of our method, we can start from a numerical
solution of the delay equation of motion, with energy $E$, and simply adjust
the two arbitrary functions $s(E,P,r)$ and $F(E,P,r)$ such that Hamiltonian (%
\ref{hamia}) has this orbit as solution. This can be done with Eqs. (\ref%
{Eq49}) and (\ref{Eq50}) and takes exactly two functions, to describe the
two particle positions as a function of the light-cone separation. The
method has a branch point at $s=0$, such that for each separation there
should be two values for the position, one in the inbound flight and another
in the outbound flight, such that we can describe twice-monotonic orbits
only. A general proof of the twice-monotonic property is given in Appendix A
of Ref. \cite{two}. At the moment the only available method to solve the
delay equations of motion is by numerical calculation. An automatic
consequence of (\ref{main3}) and (\ref{main4}) is that the orbit also
safisfies Hamitonian (\ref{hamib}).

(ii) Condition (\ref{firstwo-1}) was used as an ingredient in obtaining Eqs.
(\ref{Eq49}) and (\ref{Eq50}) of case $a$, but we never forced it directly.
An instructive direct check is to take the partial derivative respect to $E$
of the implicit energy equation ($E=H_{a}(r,p,E,P)$ ) defined by Eq. (\ref%
{hamia}), together with substitution of the potentials of Eqs. (\ref%
{solution}) and use of Eq. (\ref{condi2-2}). The result is $\partial
H_{a}/\partial E=0$, and analogously for case $b$ we obtain $\partial
H_{a}/\partial E=0$, which are the upper lines of (\ref{firstwo-1}) and (\ref%
{firstwo-2}). The lower lines of (\ref{firstwo-1}) and (\ref{firstwo-2}) are
obtained in the same way, by taking the derivatives of the implicit energy
equation respect to $P$ in each case.

(iii) After we determine $s(E,P,r)$ and $F(E,P,r)$ , substitution of $%
E=E_{o} $ and $P=P_{o}$ into the potentials produces \emph{auxiliary}
Lagrangians that have an orbit in common, precisely because of (ii). And of
course, if we did not know that yet, we would prove that this orbit
satisfies the Wheeler-Feynman equation of motion $(\delta S_{F}=0)$, as a
consequence of Eq. (\ref{consequence}).

Last, as discussed in Ref. \cite{two}, at the branch point $s=0$ and $%
r=r_{o} $ the continuity condition for the functions of Eqs. (\ref{solution}%
) and (\ref{psol}) calculates $P$ as%
\begin{equation}
P=-EF(r_{o})-\frac{e_{1}e_{2}}{r_{o}},  \label{valueP}
\end{equation}%
which will be used in the next section.

\section{\protect\bigskip Infinite-free EBK quantization and Covariance}

\bigskip\ \ Under a Lorentz transformation with boost parameter $w$ the
light-cone distance of case $a$ transforms like%
\begin{equation}
\bar{r}_{a}=\sqrt{(1+w)/(1-w)}r_{a},  \label{Lorenlight}
\end{equation}%
while in case $b$ the same formula holds with the replacement $w\rightarrow
-w.$ Equation (\ref{Lorenlight}) shows that the twice-monotonic property is
a Lorentz-invariant property, suggesting that there is no obstacle to
describe a CMF\ orbit in another Lorentz frame.

The Lorentz transformation acts on the coordinates of Hamiltonian (\ref%
{hamia}) (the $\xi _{1}$ and $\xi _{2}$ defined in Eq. ( \ref{defining}\ ) )
as a simple rescaling 
\begin{eqnarray}
\bar{\xi}_{1} &=&\frac{1}{\lambda _{a}}\xi _{1},  \label{scalecoords} \\
\bar{\xi}_{2} &=&\frac{1}{\lambda _{a}}\xi _{2}  \nonumber
\end{eqnarray}%
and also rescales the evolution parameter $\zeta $ of case $a$ to 
\begin{equation}
\bar{\zeta}=\lambda _{a}\zeta ,  \label{scaleparameter}
\end{equation}%
with $\lambda _{a}$ given by

\begin{equation}
\lambda _{a}=\sqrt{\frac{1-w}{1+w}}.  \label{deflamba}
\end{equation}%
Substituting the above Lorentz-transformed quantities into (\ref%
{Lagrangian-a}) we find another Lagrangian with the same functional form,
which proves that a CMF orbit is described in any Lorentz frame by a
Lagrangian $\bar{L}_{a}$ of the form (\ref{Lagrangian-a}). To see how the
potentials are transformed in the new frame, we make use of the explicit
solution (\ref{tablea}) in both frames, together with the condition that the
new coordinates $\bar{x}_{1a},\bar{t}_{1a},$ $\bar{x}_{1b}$ and $\bar{t}%
_{1b} $ must be the Lorentz transformed of the old coordinates $%
x_{1a},t_{1a},$ $x_{1b}$ and $t_{1b}$with the boost parameter $w$, yielding 
\begin{eqnarray}
\frac{\partial \bar{p}_{a}}{\partial E} &=&\frac{(1-w)}{(1+w)}\frac{\partial
p_{a}}{\partial E},  \label{Lorentzsymmetry} \\
\frac{\partial \bar{p}_{a}}{\partial P} &=&\frac{\partial p_{a}}{\partial P},
\nonumber
\end{eqnarray}%
where we used the differential version of Eq. (\ref{Lorenlight}) ( $d\bar{r}%
_{a}=\sqrt{(1+w)/(1-w)}dr_{a}$). Equation (\ref{Lorentzsymmetry}) shows that 
$s_{a}(r)$ as defined by Eqs. (\ref{condi2-1}) and (\ref{condi2-2}) is a
Lorentz scalar: 
\begin{equation}
\bar{s}_{a}(\bar{r}_{a},\bar{E},\bar{P})=s_{a}(\lambda _{a}\bar{r}_{a},E,P),
\label{formofs}
\end{equation}%
and that $F_{a}(r_{a})$ (as defined by Eqs. (\ref{condi2-1}) and (\ref%
{condi2-2}) ) transforms like 
\begin{equation}
\bar{F}_{a}(\bar{r}_{a},\bar{E},\bar{P})=\lambda _{a}^{2}F(\lambda _{a}\bar{r%
}_{a},E,P).  \label{formofF}
\end{equation}%
Case $b$ transforms in the same way with $\lambda _{b}=1/\lambda _{a},$
which is Eq. (\ref{deflamba}) with the substitution $w\rightarrow -w.$ This
last equation allows us to express the Hamiltonian in any frame by use of
the CMF form of $F(r)$ and a rescaling depending on the boost parameter.

It can be seen that the Lorentz transformation of Eqs. (\ref{scalecoords})\
\ is a canonical transformation that scales the momenta with the inverse
factor, $\bar{p}_{1}=\lambda _{a}p_{1}$ and $\bar{p}_{2}=\lambda _{a}p_{2}$,
such that the canonically transformed Hamiltonian is 
\begin{equation}
\bar{H}_{a}=\frac{-\lambda _{a}}{4}\{\frac{\bar{M}_{1a}^{2}(\bar{r}_{a})}{(%
\bar{p}_{1}+\frac{1}{2}\bar{V}_{1}+\frac{e^{2}}{|\bar{\xi}_{1}-\bar{\xi}_{2}|%
})}+\frac{\bar{M}_{2a}^{2}(\bar{r}_{a})}{(\bar{p}_{2}+\frac{1}{2}\bar{V}_{2}+%
\frac{e^{2}}{|\bar{\xi}_{1}-\bar{\xi}_{2}|})}\}-\lambda _{a}\bar{\phi}(\bar{r%
}_{a}),  \label{covariantH}
\end{equation}%
with\ $\bar{\phi}(\bar{r}_{a})\equiv \frac{1}{\lambda _{a}}\phi (r_{a})$, $%
\bar{V}(\bar{r}_{a})\equiv \lambda _{a}V(r_{a})$ , $\bar{M}_{1a}^{2}(\bar{r}%
_{a})\equiv M_{1a}^{2}(r_{a})$, and $\bar{M}_{2a}^{2}(\bar{r}_{a})\equiv
M_{2a}^{2}(r_{a})$. \ Notice that the new Hamiltonian $\bar{H}_{a}$ picked a
multiplicative factor of $\lambda _{a}$ , and if we perform a change to the
natural evolution parameter $\bar{\zeta}=\lambda _{a}\zeta $ of the new
Lorentz frame, it compensates exactly for that factor, going back to the
form (\ref{hamia}), such that the Hamiltonian has the same form in all
Lorentz frames (only the value of the energy is changed to $\bar{E}%
=E/\lambda _{a}$).

As the Lorentz transformation is canonical for our Hamiltonian, the action $%
I\equiv \doint pdr$ is a Lorentz-invariant, such that EBK quantization \cite%
{Percival,EBHans} is a sensible covariant quantization procedure. The need
to follow the canonical transformation by a rescaling can be avoided by use
of a Lorentz-invariant evolution parameter in the Lagrangian of Eq. (\ref%
{eqSa}). For example a Lorentz- invariant parameter can be constructed with
the help of Eqs. (\ref{formofF}) and (\ref{scaleparameter}) as 
\begin{equation}
du_{a}\equiv \frac{d\zeta }{\sqrt{F_{a}(r_{a})}}=\frac{d\bar{\zeta}}{\sqrt{%
\bar{F}_{a}(\bar{r}_{a})}}.  \label{covariantparameter}
\end{equation}%
As the momentum $p\equiv \partial L/\partial \dot{\xi}$ is independent of
parametrization, the action integral $\doint pdr$ marks the same orbits with
or without rescaling.

The action $I\equiv \doint pdr$ for the attractive problem ($e_{1}e_{2}=-1$)
can be expressed, by use of Eqs. (\ref{psol}) , (\ref{contact2}) and \ref%
{condi2-2}, as%
\begin{equation}
I=-\doint [E+\frac{1}{F(r)}(P-\frac{1}{r})]d\zeta ,  \label{action1}
\end{equation}%
and substitution of $P$ as given by Eq. (\ref{valueP}) into Eq. (\ref%
{action1}) yields 
\begin{equation}
I=\doint [\frac{1}{rF(r)}-E(1-\frac{F(r_{o})}{F(r)})]d\zeta .
\label{important}
\end{equation}%
The function $F(r_{a})$ can be calculated by elimination of $s(r)$ from Eqs.
(\ref{Eq49}) and (\ref{Eq50}) and use of (\ref{defining}) as%
\begin{equation}
F^{2}(r_{a})=\frac{(1+v_{1})}{(1-v_{1})}\frac{(1+v_{2a})}{(1-v_{2a})},
\label{Fra}
\end{equation}%
where $v_{1}$ is the velocity of particle $1$ and $v_{2a}$ is the velocity
of particle $2$ in the advanced light-cone parametrized by $r_{a}$. For a
shallow energy orbit, the particles have a low velocity in a large portion
of the orbit, such that Eq. (\ref{Fra}) predicts $F(r)\simeq 1$. The main
difference from Coulombian orbits to our covariant orbits at shallow
energies is at the collision, where Coulombian orbits reach an infinite
velocity, while the relativistic motion saturates at the speed of light.
Disregarding the collision region, substitution of $F(r)=1$ together with
the low-velocity approximation $d\zeta \simeq dt$ (valid away from the
collision), into Eq. (\ref{important}), yields exactly the one-dimensional
Coulombian WKB integral%
\begin{equation}
I=\doint \frac{dt}{r}\text{.}  \label{CoulombWKB}
\end{equation}%
Equation (\ref{CoulombWKB}) is easily derived using the Coulombian energy
and the Levi-Civita regularization, see for example the Appendix on
Levi-Civita regularization of Ref. \cite{Chaos}. At this point the reader
can see that our strange looking Hamiltonian formalism actually reduces
nicely to the Coulomb approximation for shallow orbits. The use of $I=(n+%
\frac{1}{2})\hbar $ together with the approximate formula (\ref{CoulombWKB})
predicts the well known Balmer energy terms for the one-dimensional
Coulombian two-body system. The corrections can be evaluated using the exact
covariant formula (\ref{important}) along the numerical orbits, as obtained
in Ref. \cite{Chaos} for the equal-mass case. As discussed in Ref. \cite%
{Chaos} for the equal-mass problem\cite{Chaos}, at the outbound collision $%
F(r_{a})$ is singular like $1/\sqrt{r}$, and $d\zeta =(1+v_{1})dt\simeq 
\sqrt{2}r_{a}dr_{a}$, which makes the integral of (\ref{important}) finite
(actually, for the equal-mass case the collision does not even contribute to
the integral). The merit of this procedure is that it calculates a Lamb
shift without infinities, which might prove useful for understanding the
renormalization of QED. For example, on the equal-mass two-body problem, the
orbits can be calculated numerically using the regular delay equation
developed in Ref. \cite{Chaos}. Along those \emph{regular }collision orbits,
there is nothing infinite involved in the evaluation of the action integral
of Eq. (\ref{important}) and the above covariant EBK quantization procedure
yields an electrodynamics free of infinities for the positronium atom (the
quantum energies are all finite). It would be highly desirable to generalize
this procedure to the 3-dimensional motion of the electromagnetic two-body
problem, but at present we do not know how to do that. The EBK quantization
of Wheeler-Feynman theory is discussed in \cite{EBHans,DiracHans} along with
the only physical result so far (i.e. simple EBK quantization of circular
orbits). This quantization predicts energies with finite small shifts in
agreement with the Dirac equation to order $\alpha ^{4}$ \cite{DiracHans}.
This is exactly what one expects physically, as those are circular orbits
with a finite angular momentum. The infinite Lamb shift plagues only zero
angular-momentum states, which corresponds to the orbits studied in the
present work. The possibility of using the regularization of Ref. \cite%
{Chaos} to produce an electrodynamics free of infinities for the positronium
atom suggests that further efforts to understand the regularization of the
different-mass electromagnetic two-body problem could shine light on the
renormalization of QED.

\section{\protect\bigskip Conclusions and Discussion}

\bigskip The idea to remove the field degrees of freedom from Maxwell's
electrodynamics goes back to Dirac\cite{Dirac} and later Wheeler and Feynman
planned to quantize WF2B as a means to avoid the divergencies of QED, as in
the action-at-a-distance theory the infinite number of field degrees of
freedom are absent. The difficulties in converting the Fokker Lagrangian (%
\ref{Fokker}) to Hamiltonian form have caused the famous seminar that never
came from Wheeler (see \cite{Mehra}, page 97). In chapter 5, page 97 of
reference \cite{Mehra}, Feynman says that \ ` I didn 't solve it either---a
quantum theory of half-advanced half-retarded potentials---and I worked on
it for years... '. This is still an outstanding problem today and the
difficulties in casting relativistic \emph{Lagrangian} interactions into 
\emph{Hamiltonian} form are explained in references \cite{Tretyak,Nazare}.

Our description does not violate the no-interaction theorem \cite{nointeract}%
, and there are two places where we avoid it: (i) the no-interaction theorem
is an obstacle to covariant Hamiltonian description of two interacting
particles only in fully 3-dimensional motion. We are restricted to
1-dimensional motion; (ii) Because the evolution parameter in our
Hamiltonians is not time but rather $\zeta $ for case $a$ (and $\xi $ for
case $b$), the no-interaction theorem does not apply. In principle, because
time is not the evolution parameter, even in three dimensions the
no-interaction theorem would not be an obstacle to a Hamiltonian
description, and that is an open problem. The merit of our constructive
approach is that it starts from the physical theory of action-at-a-distance
and does not involve expansions. This procedure should be compared to the
very interesting covariant extrapolation of Wheeler-Feynman theory within
constraint dynamics, which is a covariant way to introduce the correction of
order $(v/c)^{2}$ to the Coulomb problem\cite{Horace}.

The energy equation for our Hamiltonian (\ref{hamia}\bigskip ) is the
following quadratic equation for $p:$ 
\begin{equation}
p^{2}+A(E,r)p+U(E,r)=0,  \label{Schroedinger}
\end{equation}%
with%
\begin{equation}
A(E,r)=[\frac{m_{2a}^{2}-m_{1a}^{2}}{2(E+\phi )}+(V_{2}-V_{1})],
\label{phase}
\end{equation}%
and%
\begin{equation}
U(E,r)=\frac{(m_{1a}^{2}+m_{2a}^{2})}{2(E+\phi )}(P+\frac{e_{1}e_{2}}{r})+%
\frac{(m_{1a}^{2}V_{2}+m_{2a}^{2}V_{1})}{2(E+\phi )}.  \label{potential}
\end{equation}%
where $m_{1a}\equiv m_{1}+\alpha _{1}(r)$ , $m_{2a}\equiv m_{2}+\alpha
_{2}(r)$ , $V_{1}(r)$ and $V_{2}(r)$ are determined from $s(r,P,E)$ and $%
F(r,P,E)$ by Eq. (\ref{solution}). The procedure outlined above determines
the potentials only along the orbit, but it would be desirable to extend
these potentials to the whole phase with some analytical continuation and/or
use of symmetry. If \ that is accomplished, after a symmetrization, Eq. (\ref%
{Schroedinger}) can be used in a quantum canonical formalism with $p$ and $P$
being the quantum momentum operators. The trivial separation of the
center-of-mass coordinate $X$ produces a familiar looking Schroedinger
equation. This equation should be compared to the one-dimensional
Klein-Gordon equation, that has been studied in connection with the
one-dimensional hydrogen atom \cite{Spector}. The quantization of the
one-dimensional hydrogen atom interacting with the electromagnetic fields is
an interesting testbed problem, as the minimal coupling to the
electromagnetic field produces divergencies in perturbation theory that are
exactly like in the 3-dimensional problem (and which are absent in the
quantization outlined above). In that respect, Eq. (\ref{Schroedinger}) is a
properly renormalized version of the Coulomb Hamiltonian operator.

Last, our procedure could be applied to the problem of calculating level
shifts for an atom in the presence of a strong gravitational field, by
incorporating the metric tensor $g_{\mu \nu }$ in the definition of the
variables of Eq. (\ref{defining}), which should be modified to $\xi
_{1}\equiv \sqrt{g_{00}}t_{1}-\sqrt{|g_{11}|}x_{1}$ and $\zeta _{1}\equiv 
\sqrt{g_{00}}t_{1}+\sqrt{g_{11}}x_{1}.$

\section{\protect\bigskip Appendix I: Construction of the CMF frame}

For the repulsive two-electron problem ($m_{1}=m_{2}\equiv m$ and $%
e_{1}=e_{2}=e$) and along symmetric orbits [$-x_{2}(t)=x_{1}(t)\equiv x(t)$%
], the minimization of action (\ref{Fokker}) yields the following equation
of motion 
\begin{equation}
m\frac{d}{dt}(\frac{v}{\sqrt{1-v^{2}}})=\frac{e^{2}}{2r^{2}}\left( \frac{%
1-v(t-r)}{1+v(t-r)}\right) +\frac{e^{2}}{2q^{2}}\left( \frac{1+v(t+q)}{%
1-v(t+q)}\right) ,  \label{repulsemotion}
\end{equation}%
where $v(t)\equiv dx/dt$ is the velocity of the first electron, of mass $m$
and charge $e$, and the functions $r(t)$ and $q(t)$ are the time-dependent
delay and advance, defined implicitly by the light-cone conditions%
\begin{eqnarray}
r(t) &=&x(t)+x(t-r),  \label{lightcone} \\
q(t) &=&x(t)+x(t+q).  \nonumber
\end{eqnarray}%
In general, a neutral-delay equation such as Eq. (\ref{repulsemotion})
requires a pair of world-line segments as the initial condition (one
world-line segment for each particle). As discussed in Ref. \cite{wellposed}%
, the initial world-line segments can be provided in such a way that Eq.(\ref%
{repulsemotion}) is well-posed by using "maximal independent segments ". A
pair of world-line segments is called independent if the end points of each
segment lie on the forward and backward light-cones of a single point
interior to the other segment. A surprising existence theorem was proved in
Ref.(\cite{Drivergroup}) for the problem of \ Eq.(\ref{repulsemotion}) along
symmetric orbits [$-x_{2}(t)=x_{1}(t)\equiv x(t)$ ] using a Banach to Banach
contraction mapping; This theorem states that for sufficiently low energies,
Newtonian initial conditions [ $x(0)=x_{o}$ and $v(0)=v_{o}$ ] determine the
unique symmetric orbit that is globally defined (i.e., that does not runaway
at some point) \cite{Drivergroup}. \bigskip\ Later in 1981 it was proved for
this same equal- mass problem that the delay Eq. (\ref{repulsemotion} ) is
equivalent to a local Newtonian equation \cite{Zhdanov}.

The physical meaning of the above order-reduction of the delay equation is
the following: of the infinite many initial conditions necessary to solve
the delay equation of motion, only the Newtonian type "position and velocity
values at $t=0$ " suffice to determine the non-runaway orbit. The remaining
conditions either take values determined by the initial position and
velocity or the orbit will be a run-away. The above existence/uniqueness
results are certainly true in the low-velocity Coulombian limit, as the
Coulomb ODE's have unique solutions determined by initial position and
velocity. Therefore the existence/uniqueness result is either true in
general or at least up to some high energy where strange new solutions
bifurcate out. In the following we use this Newtonian order-reduction to
define the CMF frame.

We consider here the two-body problem of Eq. (\ref{Fokker}) with attractive
interaction and arbitrary masses. The construction of the Hamiltonian given
in Section II for time-reversible orbits can be generalized to arbitrary
orbits if we are able to find a Lorentz frame where this generic non-runaway
orbit is time-reversible. This frame is henceforth called the CMF frame for
the orbit. To accomplish this we start below by proving the following Lemma:

Lemma 1: Given any non-runaway orbit of Eq. (\ref{Fokker}), defined in a
given Lorentz frame, we can construct another Lorentz frame where the two
particles come to rest at the same time ($t=0$) along this orbit.

The proof will be given for the attractive case and we define the measure
time such that at $t=0$ the particles are at the origin and moving in
opposite directions with the speed of light: $x_{1}(0)=x_{2}(0)=0$, $\dot{x}%
_{1}(0)=1$ and $\dot{x}_{2}(0)=-1$ (the outbound collision). As the force is
always attractive, the absolute velocity of each particle will decrease
monotonically until it comes to rest at some turning time, and then it
accelerates again to the speed of light until the next collision, that
happens at a later time $t=T$ (the inbound collision). At this second
collision the particles bounce at each other and start again with outbound
velocities. The proof can be made as well for the case where particles pass
through each other at the inbound collision, such that the orbit is twice as
long. For simplicity we assume that they bounce, such that the unit cell of
our periodic orbit is the interval $(0,T)$, and particle one is always on
the right-hand side. The turning times of the particles are in general two
different values in the interval $(0,T)$. To fix that, we use a Lorentz
transformation with boost parameter $w$, defined such that the transformed
of events $(t_{1},x_{1}(t_{1}))$ and $(t_{2},x_{2}(t_{2}))$ be simultaneous
in the new frame. This imposes the condition%
\begin{equation}
t_{1}-wx_{1}(t_{1})=t_{2}-wx_{2}(t_{2}).  \label{cond1}
\end{equation}%
As we also want that both particles be at rest simultaneously in the new
frame, we must choose $w$ such that%
\begin{equation}
\dot{x}_{1}(t_{1})=\dot{x}_{2}(t_{2})=w.  \label{cond2}
\end{equation}%
Because the velocities are monotonic in the interval $(0,T)$, condition (\ref%
{cond2}) defines $t_{2}$ as a one-to-one function $t_{2}(t_{1})$ from the
interval $(0,T)$ onto itself. \ We can then rewrite condition (\ref{cond1})
as the zero of a function of the variable $t_{1}$ of the interval $(0,T)$ 
\begin{equation}
F(t_{1})=t_{1}-t_{2}(t_{1})-\dot{x}_{1}(t_{1})(x_{1}(t_{1})-x_{2}(t_{2}))=0.
\label{bracketing}
\end{equation}%
Using the facts that $t_{2}(t_{1}=0)=T$, $\dot{x}_{1}(0)=1$and $%
t_{2}(t_{1}=T)=0$, $\dot{x}_{1}(T)=-1$, and also that $%
x_{1}(t_{1})-x_{2}(t_{2})>0$ for all times, we find that the continuous
function $F(t_{1})$ is negative at $t_{1}=0$ and positive at $t_{1}=T$, such
that it must have a root in the interval $(0,T)$. This value of $t_{1}$
calculates the boost $w$ by Eq. (\ref{cond2}) and defines a Lorentz frame
where both particles are at rest simultaneously (the CMF), as we wanted to
demonstrate.

We have then constructed the CMF frame with the property that both particles
have zero velocity at the same time, say $t=0$, \ which we can now use as \
"Newtonian initial conditions ". As discussed above, position and velocity
initial conditions determine a unique non-runaway orbit, and since an
initial condition with both particles simultaneously at rest is invariant by
time reversal, the unique orbit determined by it must be time-reversible. We
actually need the generalization of the Banach to Banach proof of Refs. (%
\cite{Drivergroup}) for the attractive case with different masses to
complete our proof.

\end{document}